\documentstyle[12pt,epsfig]{article}
\def\ga{\mathrel{\raise.3ex\hbox{$>$\kern-.75em\lower1ex\hbox{$\sim$}}}}
\def\la{\mathrel{\raise.3ex\hbox{$<$\kern-.75em\lower1ex\hbox{$\sim$}}}}
\def\gev{{\rm \, Ge\kern-0.125em V}}
\def\tev{{\rm \, Te\kern-0.125em V}}
\def\beq{\begin{equation}}
\def\eeq{\end{equation}}

\def\ohsq{\Omega_{\chi} h^2}
\def\msn{m_{\tilde\nu}}
\def\m12{m_{1\!/2}}

\newcommand{\Zee}{$Z^0$}

\begin{document}
\begin{titlepage}
\pagestyle{empty}
\baselineskip=21pt
\rightline{hep-ph/9607292}
\rightline{CERN--TH/96-102}
\rightline{CERN--PPE/96-84}
\rightline{UMN--TH--1501/96}
\rightline{TPI--MINN--96/09}
\vskip0.65in
\begin{center}
{\large{\bf
Supersymmetric Dark Matter in the Light of LEP~1.5}}
\end{center}
\begin{center}
\vskip 0.4in
{John Ellis$^1$, Toby Falk,$^2$ Keith A.~Olive,$^2$
and Michael Schmitt$^3$
}\\
\vskip 0.2in
{\it
$^1${TH Division, CERN, Geneva, Switzerland}\\
$^2${School of Physics and Astronomy,
University of Minnesota, Minneapolis, MN 55455, USA}\\
$^3${PPE Division, CERN, Geneva, Switzerland}\\}
\vskip 0.45in
{\bf Abstract}
\end{center}
\baselineskip=18pt \noindent
We discuss the lower limit on the mass of the neutralino $\chi$
that can be obtained by combining data from $e^+ e^-$
annihilation at LEP and elsewhere with
astrophysical and theoretical considerations. Loopholes in the
purely experimental analysis of ALEPH data from the \Zee\ peak
and LEP~$1.5$, which appear when $\mu < 0$
for certain values of the sneutrino
mass $m_{\tilde \nu}$ and the ratio $\tan\beta$ of supersymmetric
Higgs vacuum expectation values, may be largely or totally
excluded by data from lower-energy $e^+ e^-$ data, the
hypothesis that most of the cosmological dark matter consists
of $\chi$ particles, and the assumption that electroweak
symmetry breaking is triggered by radiative corrections due to a
heavy top quark. The combination of these inputs imposes
$m_{\chi} \ge 21.4~\gev$, if soft supersymmetry-breaking masses
are assumed to be universal at the grand-unification scale.
\vfill
\leftline{CERN--TH/96-102}
\leftline{July 1996}
\end{titlepage}
\baselineskip=18pt
\section{Introduction}

    The recent run of LEP~at energies between $130$ and $140~\gev$,
hereafter referred to as LEP~$1.5$, has provided important new
experimental constraints on the spectrum of supersymmetric
particles~\cite{asusy,osusy,lsusy,dsusy}.
These include direct lower limits on the masses of the chargino $\chi^{\pm}$
and the right-handed selectron ${\tilde e_R}$, under certain
assumptions on the masses of other sparticles, such as the
lightest neutralino $\chi$ and the sneutrino $\tilde \nu$.
Within the context of the Minimal Supersymmetric extension of the
Standard Model (MSSM)~\cite{mssm},
these direct limits from LEP~$1.5$ have been
combined by ALEPH with previous limits on sparticle production at LEP~$1$
to obtain indirect lower limits on the neutralino mass $m_{\chi}$~\cite{achi},
which depend in particular on the assumed value of $m_{\tilde \nu}$.
Indeed, there are domains of $\tan\beta$ and $m_{\tilde \nu}$
in which $m_{\chi}$ could still vanish, in principle~\cite{achi}.

Lower limits on $m_{\chi}$ are potentially of great
interest to experimental searches for supersymmetric dark matter,
which is assumed to consist of neutralinos $\chi$~\cite{ehnos}.
Some experiments are optimized to look for relatively light
neutralinos~\cite{cresst},
and the prospective recoil energy spectrum close to
threshold is of concern to all searches~\cite{susydm}.
For these reasons, it is
useful to review the ALEPH lower limit on $m_{\chi}$~\cite{achi},
and to combine it with other phenomenological, cosmological and model
considerations, in order to evade the assumptions made
in the ALEPH analysis, and/or to strengthen the lower limit in the
presence of additional assumptions.

Cosmological arguments complementing ALEPH's purely
experimental analysis are the primary focus of this paper, whose
outline is as follows. After a brief review of
the ALEPH lower limit on $m_{\chi}$~\cite{achi},
in which we flag aspects
in which this experimental analysis may be supplemented, we
first refine some relevant phenomenological considerations. In
addition to the lower limit on $m_{\tilde \nu}$ from invisible \Zee\
decays, and the absence of sleptons at LEP~$1.5$~\cite{asusy,lsusy},
these include the interpretation of
searches at PEP and TRISTAN for single-photon events~\cite{amy},
which plays an important r\^ole in excluding a massless neutralino
in a particular domain of $\tan\beta$ and $m_{\tilde\nu}$\footnote{As
in~\cite{achi}, we assume universality of the gaugino masses:
$M_{i=1,2,3} = m_{1/2}$ at the grand-unification scale, and also universality
of the soft supersymmetry-breaking scalar masses:
$m_{\tilde \nu} = m_{\tilde e} = m_0$.}.
However, these experiments do not exclude the possibility that $m_{\tilde \nu}$
is slightly less than $m_{\chi^{\pm}}$, in which case the available lower
limits on $m_{\chi^{\pm}}$ are weakened~\cite{asusy,osusy,lsusy,dsusy}, and
the lower limit~\cite{achi} on $m_{\chi}$ requires further
discussion\footnote{We focus in this
paper on the case in which the Higgs superpotential mixing parameter
$\mu < 0$, since this is when loopholes in the ALEPH analysis~\cite{achi}
allow $m_{\chi} = 0$. We also make a few remarks
on the case $\mu > 0$, deferring a more complete analysis of
this case to a later paper.}.
For this purpose, we introduce the cosmological considerations which are
our primary interest. These include the requirement
that relic neutralinos~\cite{ehnos} not be overdense:
$\ohsq \le 0.3$, and also the preference
that they have sufficient density to be of astrophysical
interest: $\ohsq \ge 0.1$, where $h$ is the current Hubble
expansion rate, in units of $100$ km/s/Mpc. Finally, we
supplement these phenomenological constraints with the theoretical
Ansatz of electroweak symmetry breaking driven dynamically by
radiative corrections associated with a heavy top quark~\cite{ir}, which
reduces the number of supersymmetric model parameters, enabling
the lower bound on $m_{\chi}$ to be further strengthened.

In the case of $\mu < 0$ shown in Fig.~\ref{mchi_tb},
we find that the $e^+e^-$ annihilation results alone,
including the AMY analysis, enforce $m_{\chi} \ga 5\gev$, with this
limit being reached at $\tan\beta = 2$,
and cut off most of the region of low $m_\chi$ favored by
cosmology.  Combining the cosmological constraint with the
theoretical assumption of radiative electroweak symmetry breaking
strengthens significantly the lower limit on $m_{\chi}$ to $21.4~\gev$,
which is reached at $\tan\beta \simeq 1.6$, with the lower bound rising
above $51 \gev$ for $\tan\beta \ga 5$. Our bounds for $\tan\beta\ga 2.5$
are stronger than those that can be obtained indirectly from constraints
on the Higgs mass~\cite{achi} in the context of the MSSM with radiative
electroweak symmetry breaking. In the case of $\mu > 0$ (not shown), we find
for large $m_{\tilde \nu}$ that
$m_{\chi} \ga 56 \gev$ for $\tan\beta \ga 3$, which is relaxed to
$m_{\chi} \ga 36 \gev$ for $\tan\beta \la 3$.

\section{Review of Accelerator Constraints}

We first review the LEP lower limit on $m_{\chi}$ presented by the ALEPH
Collaboration~\cite{achi}, shown for $\mu < 0$ as the dashed line in
Fig.~\ref{mchi_tb}.  This results from the combination of three distinct
analyses, corresponding to lines shown in Fig.~\ref{muM2plane}:
(A)~the search for chargino pair production at LEP~$1.5$, yielding
$m_{\chi^{\pm}} \ga 67.8~\gev$ when $m_{\tilde \nu}$ is large,
(B)~the search for all channels of associated neutralino pair production at
LEP~$1.5$, and (C)~the search for $\chi\chi^\prime$ production on
the~\Zee\ peak at LEP~$1$~\cite{abig}. The latter is useful at low $\tan\beta$
when $\mu < 0$, as the high~\Zee\ statistics enable the exclusion
(C) of a wedge
of the ($M_2$, $\mu$) plane between the regions (A,B) ruled out by LEP~$1.5$,
as shown by the arrows in Figs.~\ref{muM2plane}(a,b).

Of particular interest to us are two loopholes in the ALEPH
analysis~\cite{achi} that appear when $\mu < 0$. One is that the lower
bound on $m_{\chi^{\pm}}$ is relaxed for smaller values of
$m_{\tilde\nu}$, so that the LEP~$1.5$ and LEP~$1$ constraints
no longer interlock so effectively, opening up the possibility
that $M_2 = 0$ for some values of $\mu < 0$ and $\tan\beta$
near~$\sqrt{2}$: this is indicated by the double arrow at the bottom
of Fig.~\ref{mchi_tb}. The other loophole is present only at large
$m_{\tilde\nu}$, and only for
$1 < \tan\beta < 1.02$. The first
of these loopholes arises because the lower limit on $m_{\chi^{\pm}}$
is reduced  by up to $4~\gev$ as
$m_{\tilde \nu}$ is reduced towards $m_{\chi^{\pm}}$ from above, and
then disappears entirely for $m_{\chi^{\pm}} > m_{\tilde \nu} \ga
m_{\chi^{\pm}} - 3~\gev$, in which case $\chi^{\pm}$ decay is
dominated by $\tilde \nu +$ soft lepton final states.
Bounds on chargino production reappear when
$m_{\chi^{\pm}} - m_{\tilde \nu} \ga 3~\gev$ and the lepton detection
efficiency picks up again.
The second loophole is a very small allowed region that opens up for
$\tan\beta < 1.02$ at small $M_2 \la 5~\gev$ and small negative $\mu \simeq
-30~\gev$. We note that, although $M_2=0$ is allowed in both these
loopholes, and the possibility that $\mu = 0$ could not previously be
excluded by LEP~$1$ data alone, the latter possibility has now been excluded
by the neutralino searches at LEP~$1.5$.

The ALEPH lower bound on $m_{\chi}$ based on the three types of searches
(A)-(C) mentioned above is translated into the ($\m12$, $m_0$) plane in
Fig.~\ref{m12m0plane}(a,b,c,d), as the long-dashed lines
(marked `ALEPH'). We recall that, if gaugino mass universality is assumed,
\begin{equation}
    M_2\,\simeq\,0.82\,\m12
\label{gutrelation}
\end{equation}
and that, in the limit of large $|\mu|$,
\begin{equation}
    m_{\chi}\,\simeq\,0.43\,\m12
\label{mchi}
\end{equation}
In computing the long-dashed lines in Fig.~\ref{m12m0plane}, the lower
limit on $m_{\chi}$ is obtained by varying $\mu$ over its allowed range. We see
clearly the first and more important of the two loopholes mentioned above,
where the long-dashed line in Fig.~\ref{m12m0plane} recedes to the
vertical axis.
The second and less important loophole has been ignored in drawing the
long-dashed line in Fig.~\ref{m12m0plane}(a).
One of the key issues in our analysis will
be the extent to which the two loopholes mentioned above may be plugged by
other considerations, such as the other phenomenological constraints which we
discuss next, or the cosmological and theoretical considerations which we
introduce later.

An important phenomenological constraint is the lower bound
on $m_{\tilde \nu}$ that may be inferred from the upper limit on \Zee\
decays into $\tilde \nu$ $\bar{\tilde \nu}$ imposed by the LEP~$1$
determination of the invisible \Zee\ decay width, parametrized in terms of
the equivalent number of light neutrino species, $N_{\nu}$. This is quoted by
the Particle Data Group~\cite{pdg} as yielding $m_{\tilde \nu} > 41.8~\gev$,
assuming three degenerate sneutrino species. However, this may be improved
by using the latest analysis of the~\Zee\ lineshape by the LEP Electroweak
Working Group~\cite{lepewwg}, which yields $N_{\nu} = 2.991 \pm 0.016$
corresponding to
\begin{equation}
    m_{\tilde \nu} > 43.1 \,{\rm GeV}
\label{msnu}
\end{equation}
at the $95$~\% confidence level. However, even this updated upper bound still
allows $m_{\tilde \nu}$ into the `dangerous' region where
$m_{\tilde \nu} \la m_{\chi^{\pm}}$ and the LEP~$1.5$
chargino search may lose sensitivity, so we also examine other
phenomenological constraints. For purposes of comparison, we can
express the constraint~(\ref{msnu}) in terms of the MSSM parameters
$m_0$ and $\m12$, using the standard relation
\begin{equation}
    m_{\tilde \nu}^2 = m_0^2+0.52 \m12^2-0.50 M_Z^2\: |\cos(2\beta)|
\label{slmssm}
\end{equation}
shown as the short-dashed lines in Fig.~\ref{m12m0plane}
(marked `$\tilde{\nu}$').

The ALEPH and L3 Collaborations have
also published~\cite{asusy,lsusy} lower bounds on the
${\tilde e}_R$ mass, which extend beyond the limits previously
established at LEP~$1$ by an amount that depends on the $\chi$
mass assumed. These may be
combined to give a more stringent upper limit on the cross section
for acoplanar lepton pairs.
Within the MSSM, the different slepton flavours are
almost degenerate: $m_{\tilde e} \simeq m_{\tilde \mu} \simeq
m_{\tilde \tau} = m_{\tilde \ell}$, except possibly at large
$\tan\beta$, and there is a specific relation
between the $\ell_R$ and $\ell_L$ masses:
\begin{eqnarray}
m_{{\tilde e}_L}^2 &=& m_0^2+0.52 \,\m12^2
  -0.27 M_Z^2\: |\cos(2\beta)|\cr
m_{{\tilde e}_R}^2 &=& m_0^2+0.15 \,\m12^2
  -0.23 M_Z^2\: |\cos(2\beta)|
\label{msnumssm}
\end{eqnarray}
Therefore, the limits on different slepton flavours
in different experiments may easily be combined, and
translated into the $(\m12,m_0)$ plane, as shown in Fig.~\ref{m12m0plane}
by the solid lines (marked `$\tilde{\ell}$').
For lower values of $\tan\beta$, this
combined slepton constraint improves
on the ${\tilde \nu}$ mass limit in the $\m12$ range of interest,
but still does not exclude $\chi^{\pm}$ decays into
${\tilde \nu} +$ soft lepton.

Next we consider the constraints imposed by a series of experiments looking
for single photons in $e^+ e^-$ annihilation, interpreted as $\gamma$
bremsstrahlung accompanying otherwise invisible $\nu {\bar \nu}$ or
$\chi \chi$ final states~\cite{sgamma,EH,GP}.
Experiments of this type below the~\Zee\ peak impose significant constraints
on ($m_{\chi}$, $m_{\tilde e}$) that may also be mapped into the
($\m12$, $m_0$) plane.  The AMY Collaboration has recently presented their
results for the single-photon search, and combined them with previous
measurements to exclude domains of $m_{\tilde\gamma}$ and
$m_{{\tilde e}_L}= m_{{\tilde e}_R}$~\cite{amy}.  We
have converted the AMY analysis into a constraint in the ($\m12$, $m_0$)
plane by using the correct general gaugino content of $\chi$ to
evaluate its couplings to ${\tilde e}_{L,R}$, for which we use the
mass relations (\ref{msnumssm}). We use the approximate kinematical
formulae of~\cite{EH} to adjust the AMY mass limit to take account of the
difference between the hypothetical couplings of the ${\tilde\gamma}$
and the true couplings of the $\chi$. We
show in Fig.~\ref{m12m0plane} this reinterpretation of the AMY
analysis within the MSSM. We see that this constraint excludes the
possibility that $\m12=0$, although it does not
exclude all of the region $m_{\chi^{\pm}} >
m_{\tilde \nu}$ not excluded by ALEPH.
The corresponding AMY lower limit on $m_{\chi}$ appears as the
portion of the dotted line in Fig.~\ref{mchi_tb} in the region
$1 \le \tan\beta \le 2$, excluding $m_{\chi} = 0$
in the previous loophole region indicated by the double arrow.

The small loophole for $1 < \tan\beta < 1.02$ and small $\m12, -\mu$
appears for large values of $m_0$ corresponding to
$m_{\tilde \nu} > m_{\chi^{\pm}}$.
We have checked whether this loophole
may be excluded completely by experimental upper limits on
$W^{\pm}$ decays into charginos and neutralinos. This is not
the case with the present limit of $109$ MeV on possible non-standard
contributions to the $W^{\pm}$ width, but might be the
case in the future, when improved upper limits from LEP and the
Tevatron become available. However, this small region could be
excluded completely by a small increase in the integrated
luminosity available for this analysis, as could probably be
obtained by combining the data available to the four LEP
experiments (see also~\cite{achi}).
Therefore, we believe that this small loophole
is actually illusory, though it is also excluded by the
theoretical considerations we discuss later.

\section{Cosmological Constraints}

Since one of our principal interests in this paper is supersymmetric
dark matter, we next apply the cosmological constraint that
the relic $\chi$ density not be too large, and preferably in
the range favoured by theories of structure formation
based on inflation, which predicts a total mass density
$\Omega \simeq 1$. Models with mixed hot and cold dark matter
and a flat spectrum of primordial perturbations, with a
cosmological constant and cold dark matter, and with cold dark
matter and a tilted perturbation spectrum, all favour the
range~\cite{berez}
\begin{equation}
\Omega_{\rm CDM} h^2 = 0.2 \pm 0.1
\label{omega}
\end{equation}
 We have computed $\ohsq$ in the ($M_2 $,
$\mu$) domain of interest~\cite{density}, varying $m_0$ to obtain a
result in the above range~(\ref{omega}).

The contours of $m_{\tilde \nu}$ required to obtain
the central value $\ohsq = 0.2$ for negative $\mu$ are displayed in
Fig.~\ref{muM2plane}. {\it We see that the effect of the
cosmological constraint (\ref{omega}) is, qualitatively, to constrain
the value of $m_{\tilde\nu}$, so that it is generally large and
bounded away from the dangerous loophole region}.
Figs.~\ref{muM2plane}(b,c,d) are complicated by the presence of~\Zee\ and
Higgs poles: near $m_{\chi}\approx m_Z/2$ and
$m_{\chi}\approx m_h/2$, the neutralino relic density can vary rapidly
with $m_{\chi}$, as annihilation through~\Zee\ and Higgs bosons via the
Higgsino component of the neutralino becomes enhanced. The correct
treatment of this enhancement requires calculational
techniques beyond the usual
non-relativistic expansion in powers of the annihilating relic
velocity~\cite{ehnos}, as discussed in~\cite{poles}. In
Figs.~\ref{muM2plane}(b,c,d), we label by `Z' two contours, connected
by a hatched region, corresponding to the vicinity of the~\Zee\ pole:
$\Omega_{\chi}$ first falls below 0.2
as the~\Zee\ pole is approached, and then rises back through
0.2 as the pole is left behind.  Similarly, there are two contours
labelled `H' in Figs.~\ref{muM2plane}(a,b,c).
In Fig.~\ref{muM2plane}(d), the Higgs and~\Zee\ poles for
$m_{\tilde{\nu}}=150~\gev$ and $m_{\tilde{\nu}}=200~\gev$
lie close enough together that $\Omega_{\chi}$ remains less than
0.2 for all $m_{\chi}$ between $m_h/2$ and $m_Z/2$.  In
Fig.~\ref{muM2plane}(a), there is no comparable~\Zee\ pole structure,
since for this value of $\tan\beta$
the Higgsino component of the neutralino is too small to provide
significant annihilation through~\Zee\ bosons.
The positions of the~\Zee\ and Higgs pole contours are
roughly independent of $\msn$ for large values of $|\mu|$.  However,
there are no Higgs or~\Zee\ pole contours for $\msn \la 100\gev$,
as in this case $\Omega_{\chi}$ is already
less than $0.2$ away from the poles.
Lastly, we note that stop mixing becomes significant for large $|\mu|$ and
one of the $\tilde t$ may become tachyonic when $m_{1/2}$ is small. This is
the reason why the $\msn=100\gev$ contours do not extend to $\mu=-400\gev$
in Figs.~\ref{muM2plane}(a) and 2(b). However, if one adjusts the trilinear
soft supersymmetry-breaking parameter $A_t$, these contours could be
extended to $\mu=-400\gev$ without affecting the positions of the
non-pole and~\Zee -pole contours.

The implications of this relic-density analysis are carried over to
Fig.~\ref{m12m0plane}, where we show the $(m_{1/2},m_0)$ parameter region
which admits neutralino relic densities in the cosmologically-favored range
given by (\ref{omega}).  The area below the light-shaded region leads
to relic densities $\ohsq$ which are less than 0.1 for any value of
$\mu < 0$. In the light-shaded region,
by making $|\mu|$ small (but allowed) one can produce a large Higgsino
component for the neutralino, which can then annihilate readily
through~\Zee\ and Higgs bosons. This enables $\ohsq$ to stay below
$0.3$, whatever the value of $m_0$.  For this reason,
the light-shaded region extends all the way to the top of
Fig.~\ref{m12m0plane}. Most of the band allowed by cosmology
lies above the ALEPH lower limit on $m_{1/2}$~\cite{achi}, but the
tail of it crosses the `dangerous' region where $m_{\chi^{\pm}} >
m_{\tilde \nu}$. However, we see in Fig.~\ref{m12m0plane}
that the cosmologically-allowed part of this `dangerous' region is
largely excluded by other phenomenological constraints, in particular
by the single-photon experiments~\cite{amy}. The residual part of this
`dangerous' region may be excluded by certain additional model
assumptions, as we discuss below.

\section{Electroweak Symmetry Breaking Constraints}

Further constraints on the MSSM parameters can be obtained from other,
more theoretical, considerations. In particular, it is attractive to
believe that electroweak symmetry breaking (EWSB) is driven by radiative
corrections to soft supersymmetry-breaking Higgs masses in the MSSM~\cite{ir}.
Quantifying this constraint requires additional assumptions beyond the normal
renormalization group running of the MSSM parameters, notably the assumption
that the Higgs masses are equal to the other scalar masses $m_0$ at the
supersymmetric grand unification scale. Making this universality assumption,
{\em one obtains a definite value of $\mu$}, for any given values of the other
MSSM parameters ($m_0, m_{1/2}, A, \tan\beta$)~\cite{cmssm}.  This is displayed
in the right-hand side of Fig.~\ref{muM2plane} for $m_0$ varied over the
interesting range, for a minimum value of the top mass:
$m_{\mathrm{top}}=161~\gev$.  The constraint for $\mu < 0$ is identical.

The application of the EWSB constraint on $\mu$ strengthens the ALEPH
lower bound on $m_{\chi}$. When $\mu$ is fixed in this way, as seen in
Fig.~\ref{muM2plane}, the LSP is predominantly gaugino, and the
annihilation proceeds through sfermion exchange and hence is sensitive
to $m_0$.  The dark-shaded regions in Figs.~\ref{m12m0plane}(b,c) delimit the
cosmologically-favored zones in the case $\mu<0$: the effect of the
Higgs pole~\cite{poles} is evident for $\m12 \simeq (40,\,65)~\gev$,
and the bottom side of the~\Zee\ pole appears when $\m12 \ga 90~\gev$.
The difference from the light-shaded region, which does not
exhibit such any indentations, appears because EWSB forbids one
from varying $\mu$ freely in the manner described in the previous
section. The dark regions are cut off at low $\m12$ by the absence of
radiative EWSB solutions for the values of
$m_{\rm top} \ge 161\gev$ which we assume: larger values of $m_{\rm top}$
would permit EWSB only in a smaller domain of parameter space.
The bounds derived from the chargino and neutralino searches
become more stringent when $\mu$ is fixed by EWSB, as indicated
by the solid lines in Figs.~\ref{m12m0plane}(b,c,d) (marked `EWSB').
The combination
of EWSB and cosmology shown in Fig.~\ref{m12m0plane}(b) still
allows, apparently, a tiny subset of the `dangerous' region
at small $\m12$ and $m_0 \simeq 60 \gev$. However, this subset is
in fact excluded by the LEP lower limit on the mass of the
lighter neutral scalar Higgs boson in the MSSM~\cite{MSSMHiggs},
as also discussed in~\cite{achi}.
The bounds from neutralinos cover the `dangerous' chargino region
when $\tan\beta=2$, as can be seen in Fig.~\ref{m12m0plane}(c).
For the case of
$\tan\beta=1.01$ shown in Fig.~\ref{m12m0plane}(a), we find no
experimentally consistent EWSB solution, as discussed in more
detail below. In the case of $\tan\beta=35$ shown in
Fig.~\ref{m12m0plane}(d), EWSB implies an LSP which has
a large Higgsino admixture, and whose relic density is less than 0.1
independent of $m_0$ in the range of $\m12$ plotted, so there
is no dark-shaded region in Fig.~\ref{m12m0plane}(d). We exhibit in
Figs.~\ref{m12m0plane}(b,d) how the lower limit on the chargino
mass is recovered when $m_{\tilde \nu} \la m_{\chi^{\pm}} - 3 \gev$,
once the value of $\mu$ is fixed by the EWSB constraint.

\section{Lower Bounds on the Neutralino Mass}

As seen in Fig.~\ref{m12m0plane}, for  $\mu < 0$ and
different values of $\tan\beta$, one obtains a
series of lower bounds on the neutralino mass, depending on the
assumptions applied. We exemplify these for the case
$\tan\beta = \sqrt 2$, for which LEP~$1$
alone provided no rigorous lower bound on $m_{\chi}$, whilst
the purely experimental bound from the combination of LEP~$1$ and
$1.5$ is less sensitive than for larger $\tan\beta$:
\begin{itemize}
\item
The ALEPH experimental bound for large $m_0$, which is
not a notably conservative assumption, in view of the loopholes
discussed above:
$m_{\chi}\,\ga\,17\gev$;
\item
Combination of the ALEPH experimental bound for arbitrary
$m_0$ with the limits from other unsuccessful sparticle
searches, notably the
AMY bound~\cite{amy}, which eliminates the possibility
that $M_2 = 0$:
$m_{\chi}\,\ga\,5~\gev$;
\item
Combining the above experimental limits with
the cosmological constraint on $\ohsq$, which has the effect
of favouring large values of $m_0$, and disregarding the
residual loophole region where $m_{\chi^{\pm}} \ga m_{\tilde \nu}$:
$m_{\chi}\,\ga\,16\gev$, representing a significant improvement on
the purely experimental bound in the previous case;
\item
Combining the experimental constraints with both
the cosmological constraint on $\ohsq$ and
the assumption of dynamical EWSB\footnote{We note that
applying the EWSB constraint by itself
does not improve significantly the purely experimental lower
bound on $m_{\chi}$, unless the indirect constraint
from the MSSM Higgs search is used.}, which improves further on
the bound in the previous case:
$m_{\chi}\,\ga\,24~\gev$.
\end{itemize}

\begin{figure}
\begin{center}
\mbox{\epsfig{file=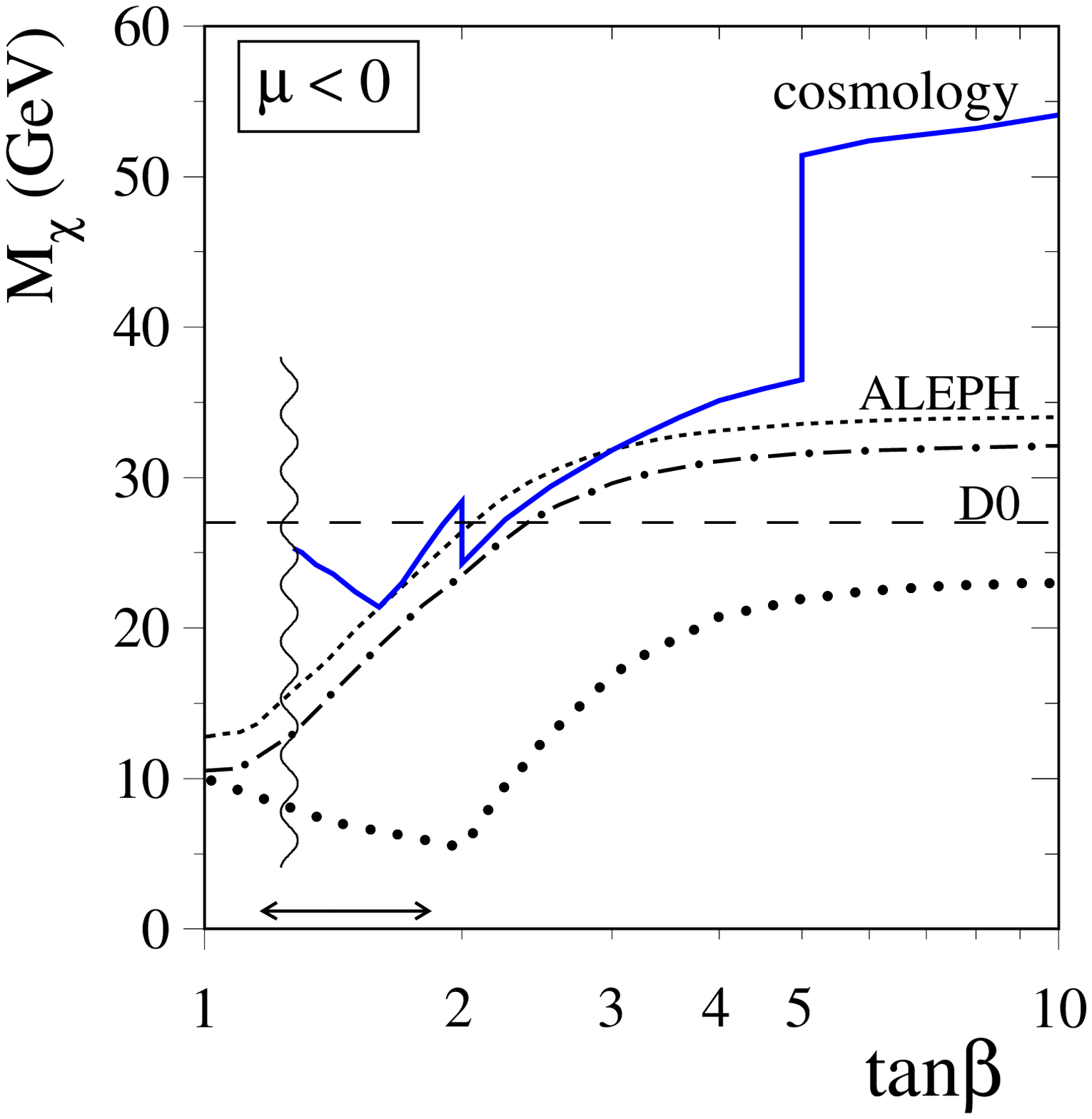,height=16.5cm%
,bbllx=10mm,bblly=55mm,bburx=170mm,bbury=215mm}}
\end{center}
\caption[.]{ \label{mchi_tb} }
\end{figure}

\begin{figure}
\begin{center}
\mbox{\epsfig{file=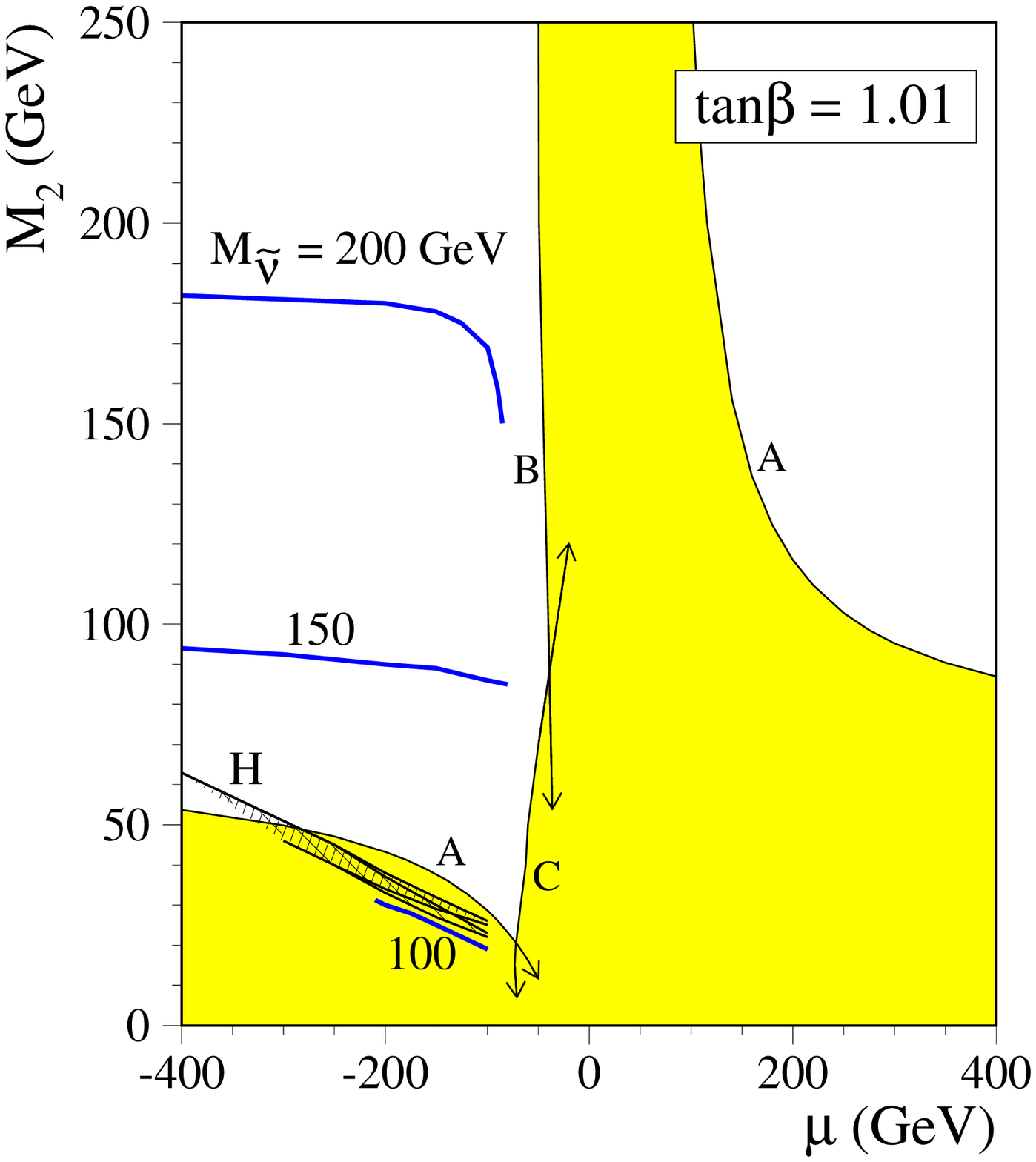,height=16.5cm%
,bbllx=15mm,bblly=15mm,bburx=165mm,bbury=205mm}}
\end{center}
\caption[.]{a \label{muM2plane} }
\end{figure}

Before discussing the dependence of these various lower bounds on $\tan\beta$,
we first note that dynamical EWSB is possible in the range of parameters
studied only if $\tan\beta$ is sufficiently large. If $\tan\beta$ is too small,
either the top quark mass $m_{\mathrm{top}} \la 160\gev$,
in conflict with the Tevatron measurements~\cite{mt},
or the running of the top-quark Yukawa coupling
becomes non-perturbative at
large scales.  Even if one believes the perturbative
renormalization-group equations in
this case, we find that the
lighter stop squark mass $m_{\tilde t}$ now falls below
the absolute lower limit from LEP. We find the restriction
\begin{equation}
  \tan\beta\,\ga\,1.2,
\label{chiewsb}
\end{equation}
which is shown as the vertical wavy line in Fig.~\ref{mchi_tb}. This
has the effect of strengthening somewhat the lower limit on
$m_{\chi}$, and also closes independently
the small $M_2, |\mu|$ loophole for $\tan\beta\,\le\,1.02$,
which, as we mentioned earlier, could also presumably be excluded
by combining the data of all the LEP collaborations.

We now return to Fig.~\ref{mchi_tb}, which displays
as functions of $\tan\beta$ our new limits on
the neutralino mass for $\mu<0$, compared with the direct ALEPH
experimental lower bound~\cite{achi} for large
$m_{\tilde \nu}$, shown as a dashed line.
The dotted line is the more conservative experimental lower
bound, obtained by allowing $m_{\tilde \nu}$ to vary freely,
but taking other sparticle searches into account.
The dash-dotted line is obtained by imposing the cosmological
constraint on $\ohsq$, and ignoring the residual loophole
region. Finally, the solid line
is obtained by combining all experimental, cosmological and EWSB
constraints.

The irregularities in this line are related to
the Higgs and~\Zee\ pole effects on $\ohsq$, and are best
illustrated by referring to Figs.~\ref{m12m0plane}(b,c).
We see in Fig.~\ref{m12m0plane}(b)
that, when $\tan\beta$ is in a range around $\sqrt{2}$,
the lower bound on $\m12$ and hence $m_{\chi}$ is given by the
intersection of the EWSB line with the central `island' of the
dark-shaded cosmological region, between the Higgs and~\Zee\
poles. As $\tan\beta$ is increased, the Higgs pole region moves
to the right, eroding the central `island', until the EWSB
line starts missing it altogether and passes along the `channel' between
the `island' and the left-hand dark-shaded region. This occurs for
$\tan\beta \simeq 1.7$, so the lowest allowed value of $m_{\chi}$ is then
given by the extreme left end of the `island' until the EWSB line
starts hitting the left-hand dark-shaded region, which occurs when
$\tan\beta \simeq 2$, as seen in Fig.~\ref{m12m0plane}(c). Thereafter, the
lower limit on $m_{\chi}$ is provided by the intersection of the EWSB line
with this left-hand region. This is the reason why the
dependence of the lower limit on $\tan\beta$ is different in the
regions $ \tan \beta = 1.2 - 1.7$, $1.7- 2$ and above $2$, as seen in
Fig.~\ref{mchi_tb}~\footnote{The lower bound on $m_{\chi}$ may be further
strengthened when $\tan\beta \la 2.5$ by taking into account the
LEP~$1$ limits on the MSSM Higgs mass, as discussed in~\cite{achi}.
The amount by which the bound is strengthened depends somewhat on
other MSSM parameters besides those discussed here, and we defer
further discussion to a subsequent paper.}.
As $\tan\beta$ is increased further, eventually the Higgs and~\Zee\
pole `channels' coalesce, and the central `island' disappears.
The contact between the EWSB line and the left-hand dark-shaded
region is lost when $\tan\beta$ reaches $\approx 5$, after which the lower
limit on $m_{\chi}$ is given by the left-hand point of the right-hand
dark-shaded region. This is the reason for the jump in the bound
at $\tan\beta \approx 5$.

We obtain from the solid line in Fig.~\ref{mchi_tb} the following
absolute lower bound on the neutralino mass for $\mu<0\,$:
\begin{equation}
m_{\chi}\,\ge\,21.4 \,\gev
\label{final}
\end{equation}
As already mentioned, we do not discuss the case $\mu > 0$ in great
detail in this paper, deferring a complete discussion to a
subsequent paper. The ALEPH analysis does not have the same sort of
loophole at large $m_0$ as in the case $\mu < 0$, and the behavior
for smaller $m_0$ is not as complex as for $\mu < 0$. A conservative
lower limit $m_{\chi} \ga 26 \gev$ is provided by the LEP~$1$ lower
limit of $45.5 \gev$ on the chargino mass, which remains valid even
when $m_{\chi^{\pm}} \ga m_{\tilde \nu}$. When $\tan\beta \ga 3$,
cosmology and EWSB strengthen this limit considerably, since $\m12$
must be to the right of a pole-induced `channel' analogous to those
shown in Figs.~\ref{m12m0plane}(b,c), and we find that
$m_{\chi} \ga 56~\gev$, for large $\msn$.
For $\tan\beta \la 3$, the limit is somewhat weaker: $m_\chi\ga36~\gev$,
again for large $\msn$.

Assuming gaugino mass universality,
our lower limit (\ref{final}) on $m_{\chi}$ can be
compared with the D0 lower limit~\cite{d0} on the gluino mass.
The D0 collaboration excludes gluino masses below $175~\gev$ for all squark
masses, with little sensitivity to $\mu$ and $\tan\beta$, and imposes a lower
limit of $220~\gev$ on degenerate gluino and squark masses. This D0 constraint
may be mapped into the ($\m12$, $m_0$) plane using the MSSM mass
relations, as indicated by the dotted lines in Fig.~\ref{m12m0plane}.
It imposes a lower limit
on $m_{1/2}$ that depends relatively weakly on $m_0$, and is
stronger than that of LEP~$1$ and~$1.5$ for small
values of $\tan\beta$, as may be seen
from the horizontal long-dashed line in Fig.~\ref{mchi_tb}.
However, Fig.~\ref{mchi_tb} shows that
stronger limits on $m_{\chi}$ may be obtained from
cosmological and/or dynamical EWSB considerations\footnote{We
emphasize that the D0 constraint shown in Figs.~\ref{mchi_tb},~\ref{m12m0plane}
holds only if one imposes equality between the gluino and
electroweak gaugino masses at the GUT scale.
It is clear from Fig.~\ref{m12m0plane} that the purely experimental LEP~$1.5$
constraint on $m_{1/2}$ at small $\tan\beta$ could
also become stronger than the limit
inferred from D0 if the GUT relation
were violated by a factor of two or so. However, an exploration
of non-universality in the gaugino masses lies beyond the scope
of this paper.}.

\section{Conclusions and Prospects}
\par
The recent ALEPH experimental analysis~\cite{achi} has enabled a
qualitative step forward to be made in bounding the mass of
the lightest neutralino. The LEP~$1$ and~$1.5$ data combine to
exclude regions of parameter space that could not be
excluded by either data set in isolation. However, the ALEPH
analysis could not exclude entirely the possibility
when $\mu < 0$ that
$\m12\,=\,0$, and hence $m_{\chi}\,=\,0$. We have shown in
this paper that this loophole may be excluded by other
experimental data~\cite{amy}, leading to a lower bound
on $m_{\chi}$ that
may be strengthened significantly by constraining the relic
density $\ohsq$, with further improvement possible if one
assumes dynamical EWSB. Combined,
these constraints provide a lower limit on $m_{\chi}$ that
is stronger even than that inferred indirectly from the
unsuccessful gluino search of the D0 collaboration. In the
near future, data from higher-energy runs of LEP around and
above the $W^+ W^-$ threshold (which we call LEP~$2W$) will be able
to explore definitively the loopholes in the recent ALEPH
experimental analysis, making the theoretical appeal to
cosmology and EWSB unnecessary.

Although we do not wish to
prejudge the results of these future searches,
it already seems likely that the neutralino mass must be
considerably larger than previous experimental limits,
with strong implications for some dark matter search experiments.
Some of these are optimized for $m_{\chi} \la 10\gev$~\cite{cresst},
whilst all direct search experiments
may benefit from the higher nuclear recoil energies now to be
expected. On the other hand, rates for both direct and indirect
dark matter searches are generally reduced as $m_{\chi}$
increases. Thus, the recent LEP~$1.5$ run and future
LEP~$2W$ data at
higher energies may have significant impact on the search for
supersymmetric dark matter.

\vskip 0.5in
\vbox{
\noindent{ {\bf Acknowledgments} } \\
\noindent  J.E. would like to thank the University of Minnesota,
Lawrence Berkeley National Laboratory and the
Berkeley Center for Particle Astrophysics
for kind hospitality while parts of this work were being done.
M.S. would like to thank Jean-Francois Grivaz and Laurent Duflot
for valuable discussions.
This work was supported in part by DOE grant DE--FG02--94ER--40823.}

\newpage
\noindent{\bf{Figure Captions}}

\vskip.2truein

\begin{itemize}
 \item[]
\begin{enumerate}

\item[Fig.~1)]The ALEPH lower limit on $m_{\chi}$~\cite{achi}
for $\mu < 0$ and for large $m_{\tilde \nu}$
(short-dashed line) is compared, as a function of $\tan\beta$,
with the results obtained in the text by making different
phenomenological and theoretical inputs. The dotted line is
obtained by combining the AMY constraint~\cite{amy} with other
unsuccessful searches for sleptons and sneutrinos:
it excludes the region of $\tan\beta$, indicated by a double arrow,
where the ALEPH experimental limit does not exclude $m_{\chi} = 0$.
The dash-dotted line is obtained by requiring also that the cosmological
relic neutralino density fall within the preferred range~(\ref{omega}).
The solid line is obtained by combining these experimental and
cosmological inputs with the assumption~\cite{ir} of dynamical
electroweak symmetry breaking. The vertical wavy line
indicates the lower limit on $\tan\beta$ in such dynamical electroweak
symmetry breaking models. The horizontal long-dashed line is that obtained
from the D0 gluino search~\cite{d0}, assuming gaugino mass universality.

\item[Fig.~2)] The region of the ($\mu$, $M_2$) plane excluded by direct
searches~\cite{asusy} for (A)~charginos at LEP~$1.5$, (B)~neutralinos at
LEP~$1.5$ and (C)~\Zee\ decays into $\chi \chi^\prime$ at LEP~$1$ for
$\tan\beta = 1.01$, $\sqrt{2}$, $2$, $35$ are indicated in~(a,b,c,d)
respectively by thin solid lines.
Contours of $m_{\tilde \nu}$ (in~GeV) required in the MSSM
to obtain $\Omega_{\chi} h^2 = 0.2$ for $\mu < 0$ are indicated by
thick solid lines. The hatched
regions indicate where the Higgs and~\Zee\ poles suppress the relic
density, as discussed in the text. Values of $\mu$ required by dynamical
electroweak symmetry breaking for the indicated
values of $m_0$ (in~GeV) are shown as short-dashed
lines in (b,c,d) for $\mu > 0$: identical values would be required for
$\mu < 0$.

\item[Fig.~3)]For $\tan\beta = 1.01$~(a), $\sqrt{2}$~(b), $2$~(c)
  and $35$~(d), we display for $\mu < 0$ the domains of the ($m_{1/2}$,$m_0$)
  plane that are excluded by the ALEPH chargino and neutralino
  searches~\cite{achi} (long-dashed line), by the limit~(\ref{msnu}) on
  $m_{\tilde \nu}$ (short-dashed line), by the LEP limits~\cite{asusy,lsusy}
  on slepton production (solid line), by single-photon measurements~\cite{amy}
  (grey line), and by the D0 limit on the gluino mass~\cite{d0} (dotted line).
  The region of the plane in which $0.1 < \Omega_{\chi} h^2< 0.3$ for some
  experimentally-allowed value of $\mu<0$ is light-shaded, and the region of
  the plane in which $0.1 < \Omega_{\chi} h^2< 0.3$ for $\mu$ determined by
  dynamical electroweak symmetry breaking is shown dark-shaded in~(b,c).
  The constraint derived from the ALEPH searches when imposing dynamical
  electroweak symmetry breaking~(EWSB) is also shown as a solid line
  in~(b,c,d).
\end{enumerate}
\end{itemize}
\newpage
%
%
\setcounter{figure}{1}
\begin{figure}
\begin{center}
\mbox{\epsfig{file=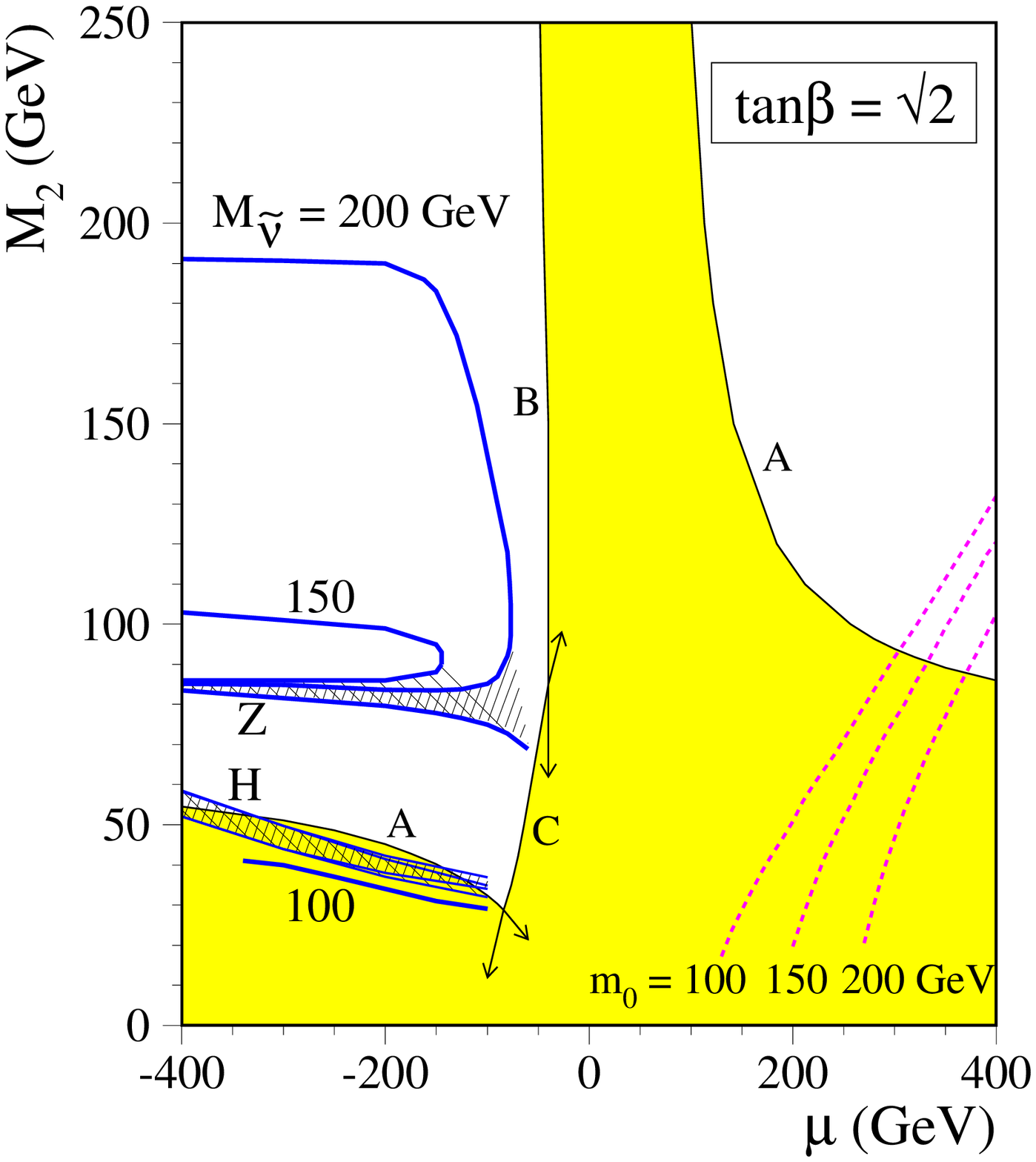,height=16.5cm%
,bbllx=15mm,bblly=15mm,bburx=165mm,bbury=205mm}}
\end{center}
\caption[.]{b }
\end{figure}
\setcounter{figure}{1}
\begin{figure}
\begin{center}
\mbox{\epsfig{file=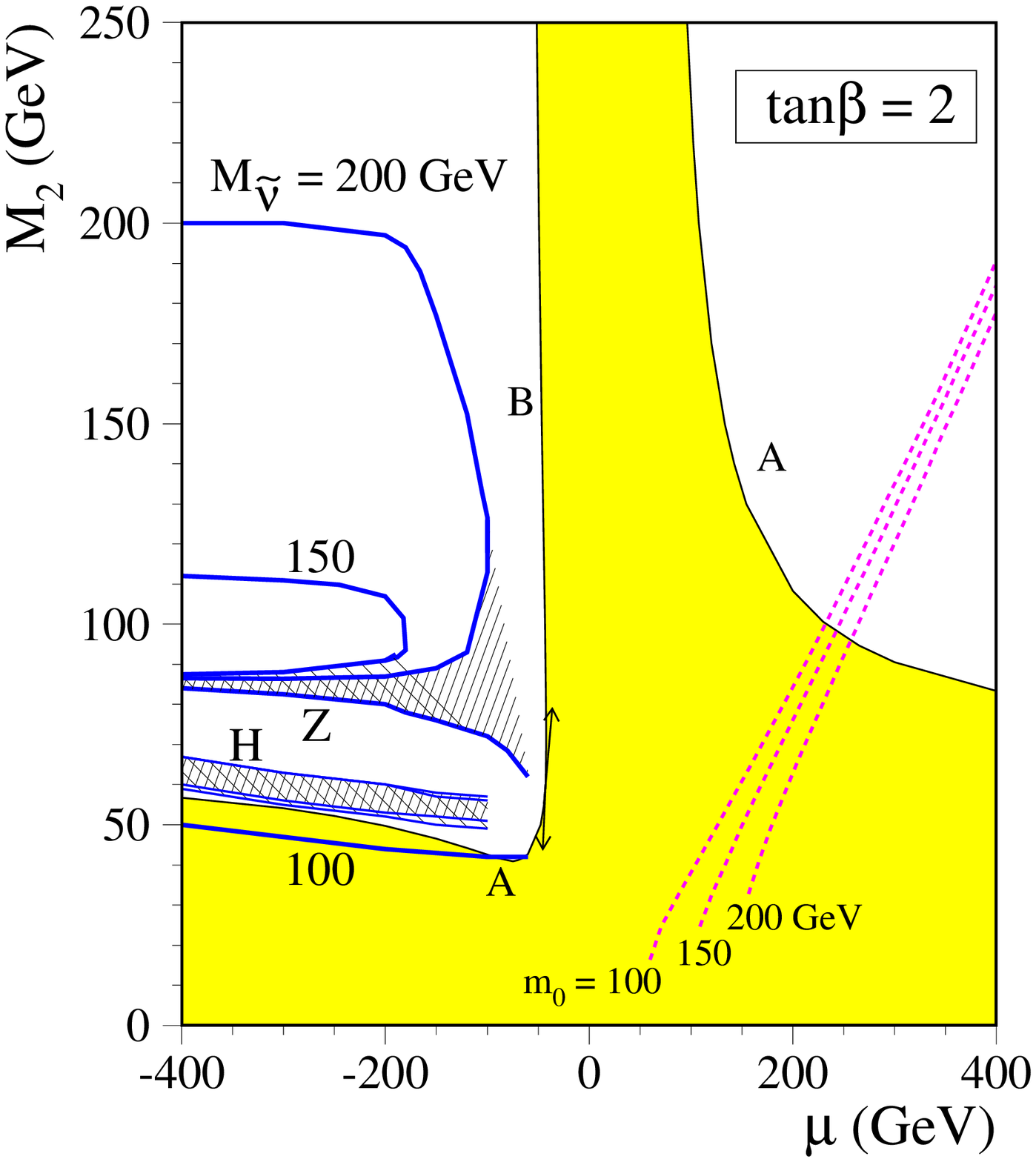,height=16.5cm%
,bbllx=15mm,bblly=15mm,bburx=165mm,bbury=205mm}}
\end{center}
\caption[.]{c }
\end{figure}
\setcounter{figure}{1}
\begin{figure}
\begin{center}
\mbox{\epsfig{file=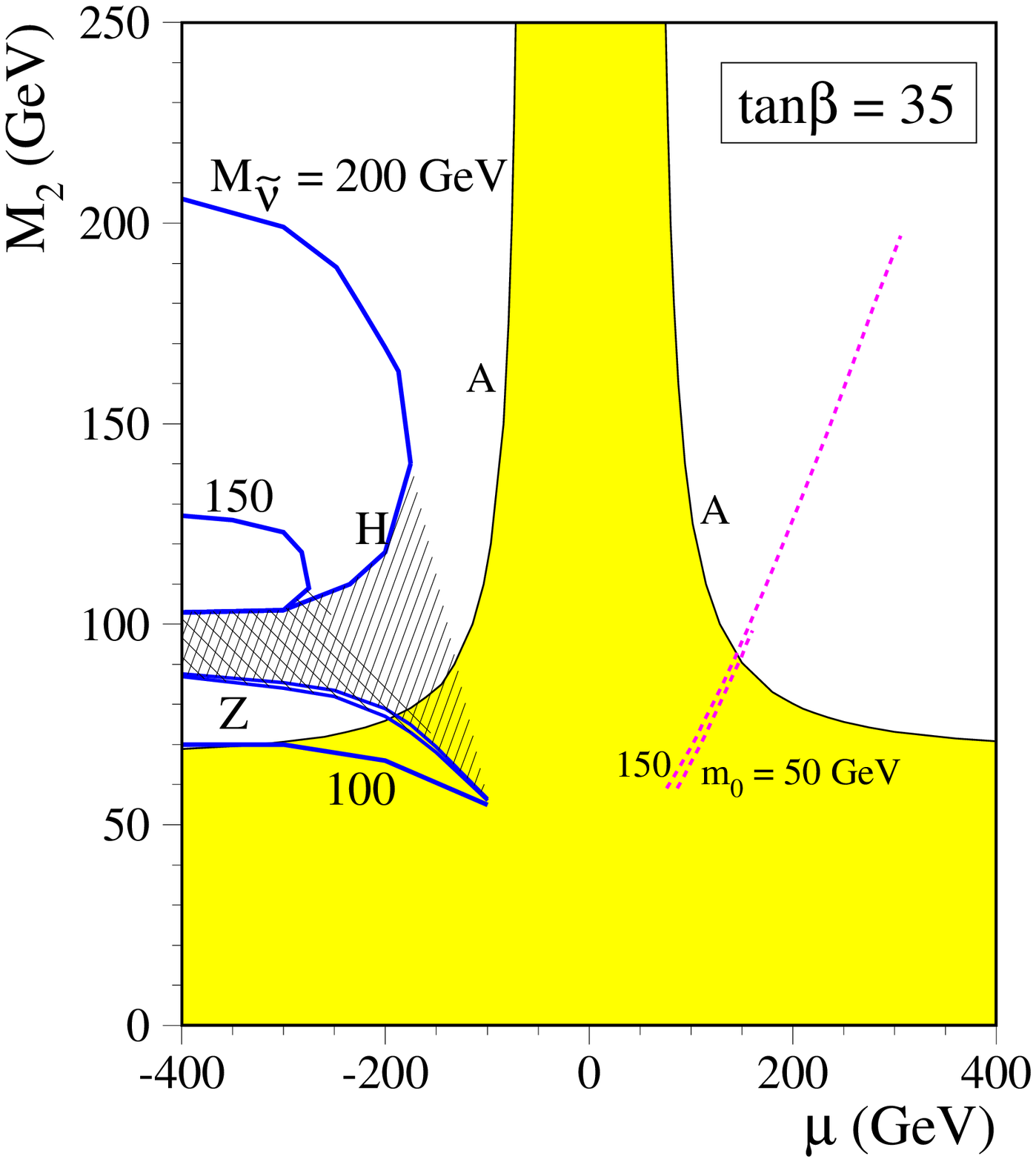,height=16.5cm%
,bbllx=15mm,bblly=15mm,bburx=165mm,bbury=205mm}}
\end{center}
\caption[.]{d }
\end{figure}
\setcounter{figure}{2}
\begin{figure}
\begin{center}
\mbox{\epsfig{file=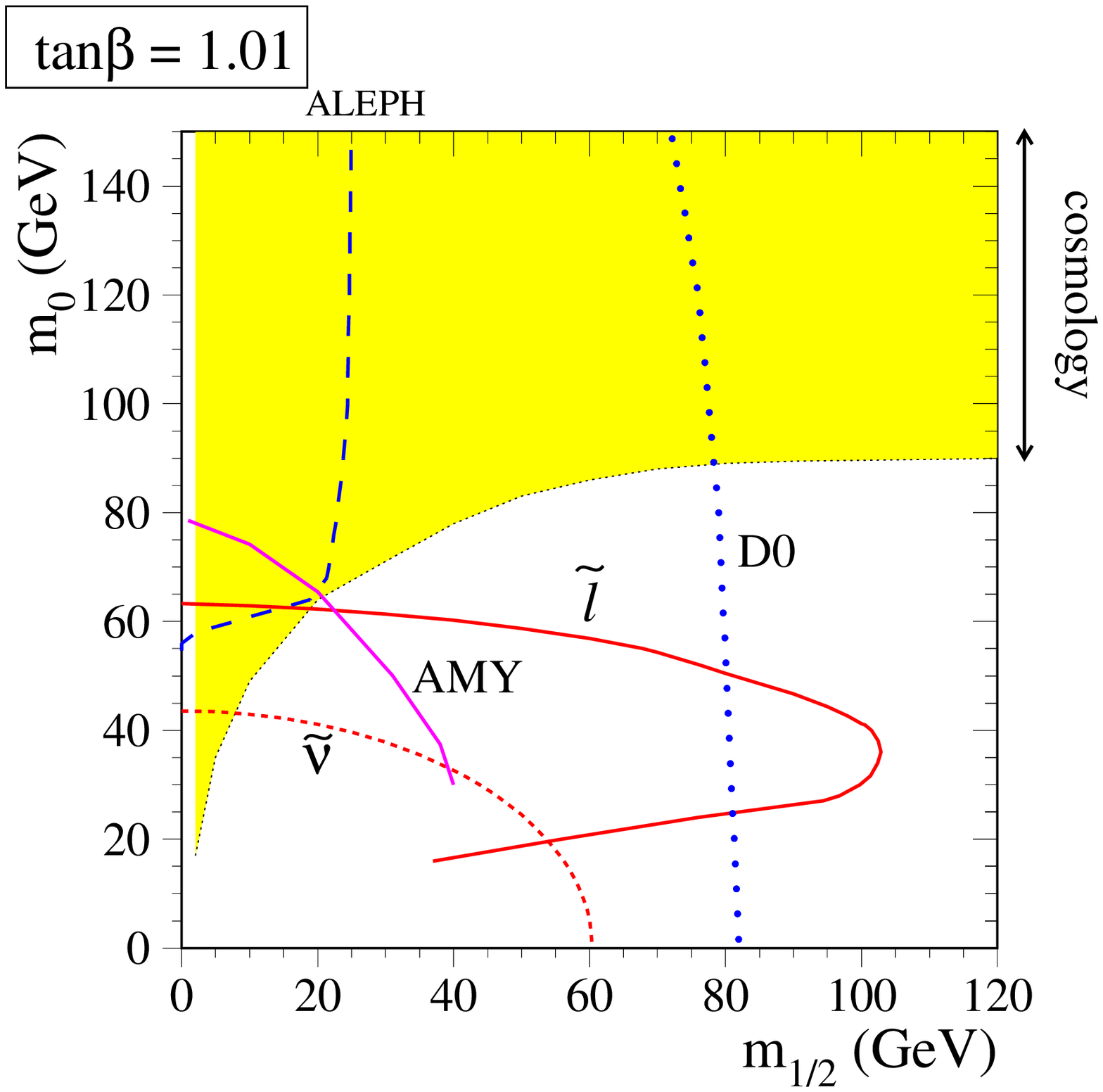,height=16.5cm%
,bbllx=12mm,bblly=30mm,bburx=160mm,bbury=235mm}}
\end{center}
\caption[.]{a \label{m12m0plane} }
\end{figure}
\setcounter{figure}{2}
\begin{figure}
\begin{center}
\mbox{\epsfig{file=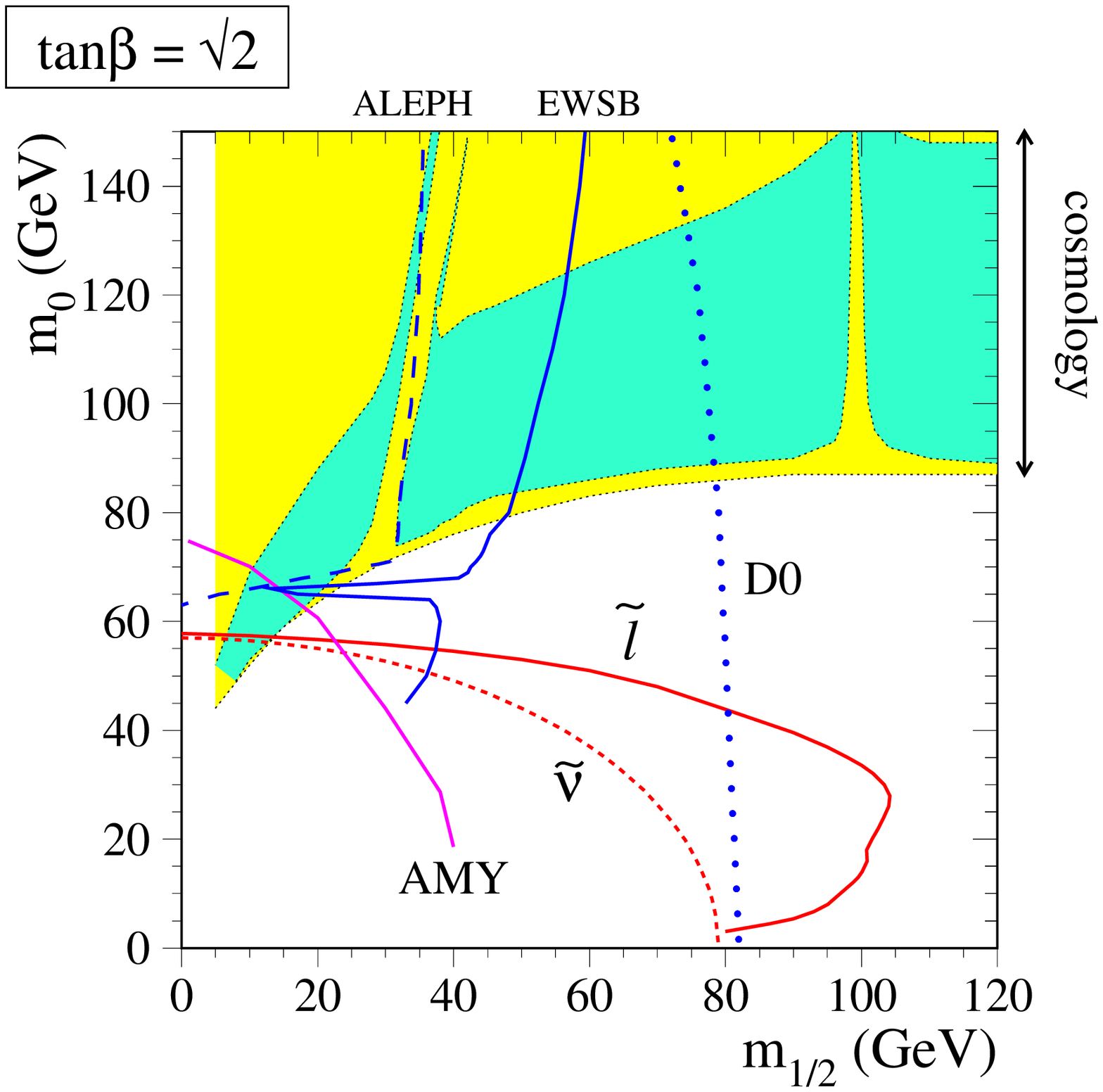,height=16.5cm%
,bbllx=12mm,bblly=30mm,bburx=160mm,bbury=235mm}}
\end{center}
\caption[.]{b }
\end{figure}
\setcounter{figure}{2}
\begin{figure}
\begin{center}
\mbox{\epsfig{file=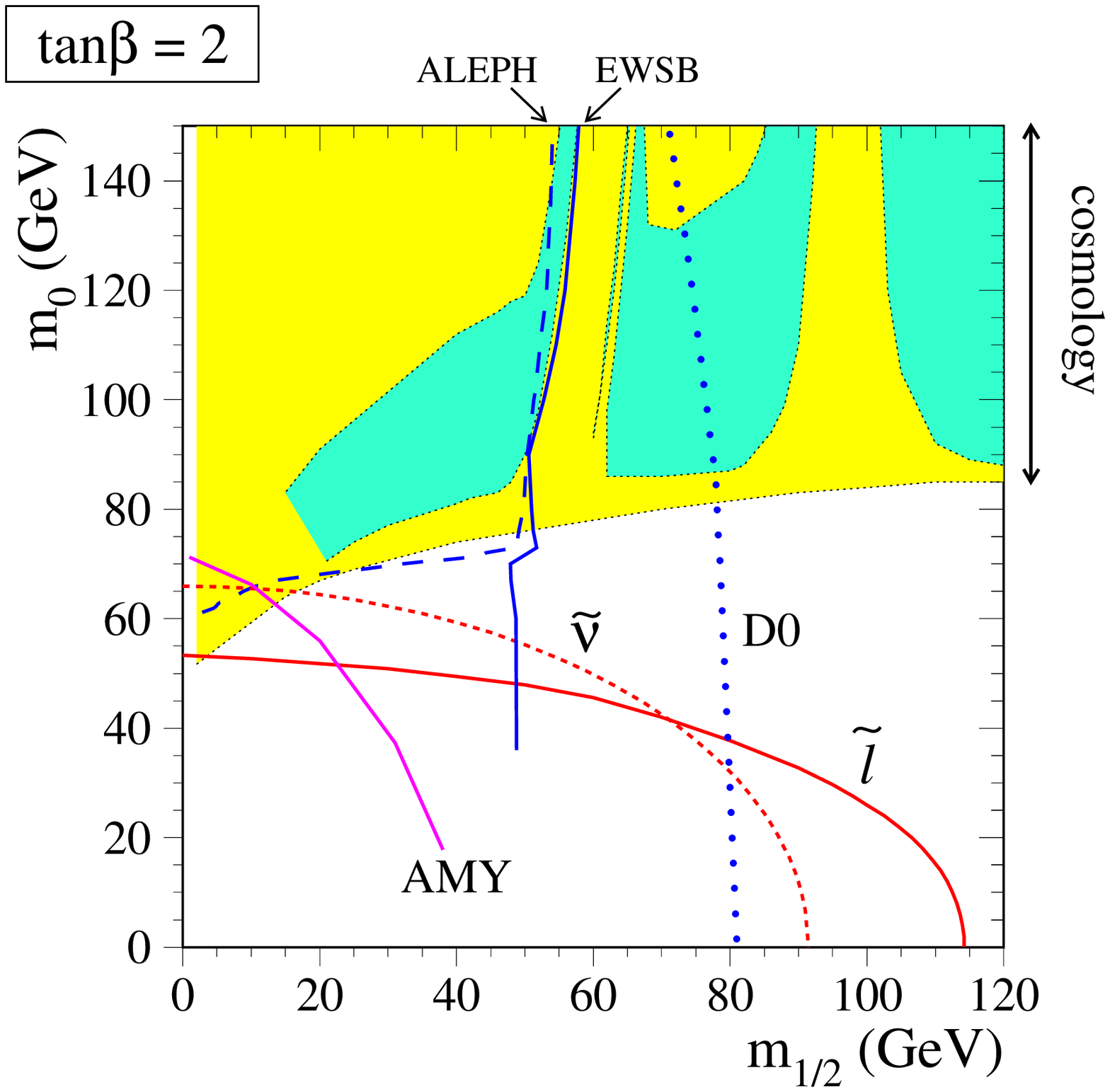,height=16.5cm%
,bbllx=12mm,bblly=30mm,bburx=160mm,bbury=235mm}}
\end{center}
\caption[.]{c }
\end{figure}
\setcounter{figure}{2}
\begin{figure}
\begin{center}
\mbox{\epsfig{file=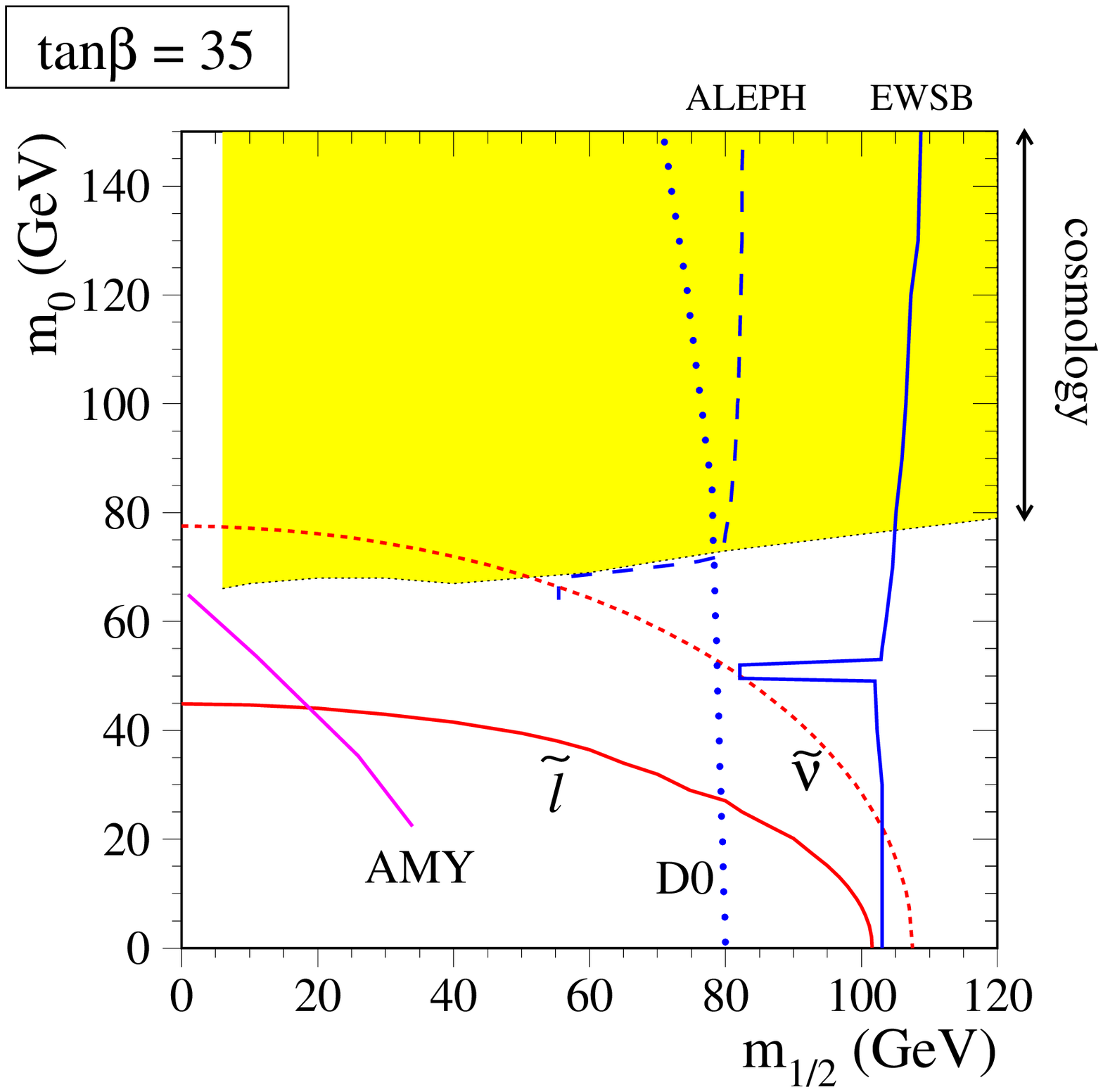,height=16.5cm%
,bbllx=12mm,bblly=30mm,bburx=160mm,bbury=235mm}}
\end{center}
\caption[.]{d }
\end{figure}
%
\end{document}